# Sensitivity of Fitting High Order Zernikes

Gregg M. Gallatin
AppliedMathSolutions.com
gregg@appliedmathsolutions.com

## I. ABSTRACT

The sensitivity of fitting Zernike coefficients to wavefront data due to an error or uncertainty in the exact position of the edge of the wavefront data is derived. The results indicate fitting higher and higher order Zernikes rapidly becomes a useless exercise unless the true edge of the wavefront data is known with sufficient precision.

## II. INTRODUCTION

The Zernike functions, usually referred to simply as "Zernikes" [1] are a complete set of orthogonal basis functions defined on the unit circle. Wavefronts in a optical system are generally circular, and hence the Zernikes are commonly used to represent or analyze wavefront data, i.e. optical aberrations. The Zernikes are polynomials in the radial direction, $\rho$, and sines and cosines in the tangential or angular direction, $\varphi$. Historically only low order Zernikes have been used for representing wavefronts, i.e., those which depend on the radius up to around the 10th or 12th power. In part, this is because the low order Zernikes directly characterize the most important optical properties of a given optical system, but also in part due to historical limitations of measuring and analyzing wavefront data.

In order to represent the wavefront as a linear superpostion of Zernikes, the radius of the wavefront data must be normalized to live on the unit circle. i.e., $r = \rho/a$ where $a$ is the maximum value of $\rho$. $\rho$ can in units of distance or spatial frequency or simply pixels. For example, in terms of spatial frequency at a given wavelength $\lambda$ and Numerical Aperture, $NA$, $a = NAk$ with $k = 2\pi/\lambda$. Because of this need to normalize the radius, any error or uncertainty in the value of $a$ used to do the normalization will impact the fitted results.

I have heard people talk about fitting wavefront data to much higher order Zernikes than has been done in the past. Such high order Zernikes of course depend on higher and higher powers of the radius, e.g., $r$ to the 20th power or more. This raises an interesting question. Because the Zernikes are polynomials in $r$, the higher the order, the larger will be the absolute value of their radial slope as $r$ approaches 1. This increase in the slope increases the sensitivity of the higher order fitted coefficient values to variations or errors or uncertainties in the value of $a$ used to do the fitting. Here, we evaluate this sensitivity analytically as a function of the order. The results given below show that the increase in the sensitivity with increasing order renders the fitted coefficient values virtually meaningless, depending on the required accuracy, above a given order for a given error or uncertainty in the value of $a$. This error or uncertainty in the value of $a$ also breaks the orthogonality of the Zernikes leading to non-zero cross coefficient values. To put it succinctly: Fitting wavefront data to higher and higher order Zernikes rapidly becomes, from a practical standpoint, pointless.

What is commonly done instead of using higher order Zernikes, is to fit to a fixed number of lower order Zernikes, subtract the fitted wavefront from the wavefront data and analyze the remaining or residual wavefront data using Fourier Transforms. By definition, the residual wavefront data contains spatial frequencies higher than those in the Zernikes used to do the fit. The Fourier amplitudes obtained this way are further reduced to an annular Power Spectral Density (PSD) by squaring and averaging over $\varphi$ at each spatial frequency. The resulting PSD is directly useful optically since it corresponds to the spatial distribution of the intensity of "scattered" light surrounding the nominal point spread function in an imaging system, i.e, it gives the "halo" of light stretching beyond the point spread function [2] . This halo is commonly referred to as "flare" [3] .

As might be expected, the same conclusion applies to the sensitivity of fitted coefficients when using any set of radial polynomial basis functions, defined on the unit circle. Certainly other issues of accuracy arise when dealing with high order polynomials, such as the numerical issues discussed by Forbes [4] and those discussed by Will and Fienup [5] .

## III. BRIEF REVIEW OF THE ZERNIKES

In polar coordinates on the unit circle, $r = 0$ to 1 and $\varphi = 0$ to $2\pi$, the Zernikes are defined by [1]

$$Z_n^{\pm m}(r,\varphi) = \begin{cases} R_n^m(r)\cos(m\varphi) \\ R_n^m(r)\sin(m\varphi) \end{cases} \tag{1}$$

where $+m$ corresponds to the $\cos(m\varphi)$ terms and $-m$ to the $\sin(m\varphi)$ terms. Here $n$ and $m$ are non-negative integers used to label the different Zernike functions: $n = 0, 1, 2, \cdots$ and $0 \leq m \leq n$ with $n - m$ even.

The radial functions $R_n^m(r)$ are defined by

$$R_n^m(r) = \begin{cases} \sum_{k=0}^{(n-m)/2} \frac{(-1)^k (n-k)!}{k!\left(\frac{n+m}{2}-k\right)!\left(\frac{n-m}{2}-k\right)!} r^{n-2k} \text{ for } n-m = \text{even} \\ 0 \text{ for } n-m = \text{odd} \end{cases} \tag{2}$$

The $Z_n^{\pm m}(r,\varphi)$, as defined above, are orthogonal on the unit circle, i.e.,

$$\frac{1}{\pi}\int_0^1 rdr \int_0^{2\pi} d\varphi Z_n^{\pm m}(r,\varphi) Z_{n'}^{\pm m'}(r,\varphi) = \frac{1}{2(n+1)}\delta_{n,n'}\delta_{m,m'} \tag{3}$$

Here $\delta_{a,b} = 1$ for $a = b$ and is zero otherwise. The so called "rms normalization" factors are, by definition, given by $\sqrt{2(n+1)}$. Without this normalization the Zernikes are termed "fringe normalized" since at the "fringe" or edge of the circle, $r = 1$, we have

$$R_n^m(1) = 1 \tag{4}$$

for all $n$ and $m$ (with $n - m =$ even).

Commonly the Zernikes are "reordered" to depend on just one parameter, $n = 1, 2, 3$,instead of the original two with $m \to m(n)$ ranging from 0 to $n$.

$$Z_n^m(r,\varphi) \to Z_n(r,\varphi) = R_n^{m(n)}(r)\left(\sin[m(n)\varphi] \text{ or } \cos[m(n)\varphi]\right) \tag{5}$$

The $Z_n(r,\varphi)$ defined this way are also fringe normalized and orthogonal with respect to $n$, and the sine or the cosine are both used for each value of $m(n)$. Historically various reorderings have been used. But generally $Z_1$ up to roughly $Z_{37}$ have the same reordering in most cases [6].

## IV. DERIVATION

Because the Zernikes form a complete set of basis functions, any function defined on the unit circle can be represented as a linear superposition of Zernikes. (There are some limits to this which are, for the most part, only of interest to mathematicians.) Wavefronts are usually circular in optical systems, hence the utility of the Zernikes in representing wavefront data. An analytical or numerical representation of a wavefront, $w(\rho,\varphi)$ with $0 \leq \rho \leq a_w$ can be mapped to the unit circle by letting $r = \rho/a_w$. Note that, depending on the particular case, $\rho$, and hence also $a_w$, can be in distance units or spatial frequency units or in pixel number.

The fit of such a wavefront to the Zernikes, up to $n = N$ is then written

$$\begin{aligned} w(\rho,\varphi) &= \sum_{n'=0}^{N}\sum_{m'=0}^{n'}\left(c_{n',+m'}Z_{n'}^{+m'}\left(\frac{\rho}{a_w},\varphi\right) + c_{n',-m'}Z_{n'}^{-m'}\left(\frac{\rho}{a_w},\varphi\right)\right) \\ &= \sum_{n'=0}^{N}\sum_{m'=0}^{n} c_{n',\pm m'}Z_{n'}^{\pm m'}\left(\frac{\rho}{a_w},\varphi\right) \end{aligned} \tag{6}$$

Remember that $Z_{n'}^{\pm m'}\left(\frac{\rho}{a_w},\varphi\right)$ is, by definition, zero unless $n' - m'$ is even.

The fitted Zernike coefficients, $c_{n,\pm m}^{\text{fit}}$, can be determined, nominally, using the orthonormality condition with a fitting radius $a_{\text{fit}}$ by setting $r = \rho/a_{\text{fit}}$ and computing

$$c_{n,\pm m}^{\text{fit}} = \frac{1}{a_{\text{fit}}^2}\int_0^{a_{\text{fit}}} d\rho\rho \int_0^{2\pi} d\varphi w(\rho,\varphi) Z_n^{\pm m}\left(\frac{\rho}{a_{\text{fit}}},\varphi\right)$$

Using the orthogonality of the sine and cosine factors the Zernikes, this reduces to

$$\begin{aligned} c_{n,\pm m}^{\text{fit}} &= \frac{2(n+1)}{\pi} \sum_{n'=1}^{N} c_{n',\pm m'} \int_0^{a_{\text{fit}}} \frac{d\rho\rho}{a_{\text{fit}}^2} \int_0^{2\pi} d\varphi Z_{n'}^{\pm m'}\left(\frac{\rho}{a_w},\varphi\right) Z_n^{\pm m}\left(\frac{\rho}{a_{\text{fit}}},\varphi\right) \\ &= 2(n+1) \sum_{n'=0}^{N} c_{n',\pm m} \int_0^{a_{\text{fit}}} d\rho\rho R_{n'}^{\pm m}\left(\frac{\rho}{a_w}\right) R_n^{\pm m}\left(\frac{\rho}{a_{\text{fit}}}\right) \\ &= 2(n+1) \sum_{n'=0}^{N} c_{n',\pm m} \int_0^{1} drr R_{n'}^{\pm m}\left(\frac{a_{\text{fit}}}{a_w}r\right) R_n^{\pm m}(r) \end{aligned} \tag{7}$$

Obviously as long as $a_w = a_{\text{fit}}$ everything works as it should and there is no issue. Of course there can be other numerical issues which depend on how the integrals are represented numerically or on how a least squares fit is used to find the coefficients, but neither of these is the point here since we will evaluate Eq (7) analytically for the case where $a_w \neq a_{\text{fit}}$.

Two comments before continuing: First, how can we have $a_w \neq a_{\text{fit}}$? Interferograms of the pupil in an optical system are generally captured on an array of detectors, e.g., a CCD camera. Hence the wavefront data which is captured in the interferogram is distributed spatially across the pixels of the detector array. The precise edge of the wavefront is then at some fraction of a pixel around the edge of the data. There is no absolute method for determining precisely where the edge lies in any given pixel. Second, consider the reason why the absolute value of the slope of a Zernike, $|\partial_r R_n^m(r)|$, increases with increasing order as $r \to 1$. Certainly the single term $r^p$ has increasing slope as $r \to 1$. But the $R_n^m(r)$ are polynomials and so any lower power term will, depending on coefficients, dominate any higher power term in the region $r \lesssim 1$. But, as defined, the $R_n^m(r)$ have approximately $n/2$, more or less, evenly spaced zeros between $r = 0$ and $r = 1$, hence the higher the order the more zero crossings, with the last zero crossing getting closer and closer to $r = 1$ as the order increases. By definition, $R_n^m(1) = 1$ and so $|\partial_r R_n^m(r)|$ must increase with increasing order since it must go from 0 to 1 in a shorter and shorter distance.

To see slope dependence of $c_{n,\pm m}^{\text{fit}}$ on the difference between $a_w$ and $a_{\text{fit}}$ along with the breakdown of orthogonality, let $a_{\text{fit}} = a_w + \delta a_w$ and Taylor expand Eq (7) in powers of $\delta a_w$

$$\begin{aligned} c_{n,\pm m}^{\text{fit}} &= 2(n+1) \sum_{n'=0}^{N} c_{n',\pm m} \int_0^1 drr R_{n'}^{\pm m}\left(\left(1+\frac{\delta a_w}{a_w}\right)r\right) R_n^{\pm m}(r) \\ &= 2(n+1) \sum_{n'=0}^{N} c_{n',\pm m} \int_0^1 drr \left(R_{n'}^{\pm m}(r) + \left(\partial_r R_n^{\pm m}(r)\right)\frac{\delta a_w}{a_w} r + \cdots\right) R_n^{\pm m}(r) \\ &= c_{n,\pm m} + 2(n+1) \sum_{n'=0}^{N} c_{n',\pm m} \int_0^1 drr \left(\partial_r R_n^{\pm m}(r)\right)\frac{\delta a_w}{a_w} r R_n^{\pm m}(r) + \cdots \end{aligned} \tag{8}$$

The second term, and higher terms, above explicitly show the breakdown of orthogonality along with the dependence of $c_n^{\text{fit}}$ on the product $r^2 \times (\partial_r R_n^m(r)) \times R_n^m(r)$, all three factors of which rapidly increase as $r \to 1$. For small values of $n$ the second term above is a sufficient approximation since it is nominally small compared to $c_{n,\pm m}$ itself. But since the $R_n^m(r)$ are polynomials of $n^{th}$ order in $r$, it follows that the second term above, by itself, is not a good approximation for large $n$. This polynomial dependence also means that the overlap integrals above can be done exactly.

To evaluate $c_{n,\pm m}^{\text{fit}}$ analytically, first simplify notation by setting

$$A_{n,m,k} = \frac{(-1)^k (n-k)!}{k!\left(\frac{n+m}{2}-k\right)!\left(\frac{n-m}{2}-k\right)!} \tag{9}$$

which gives

$$
\begin{aligned}
c^{\mathrm{fit}}_{n,\pm m} &= 2\left(n+1\right)\sum_{n'=0}^{N} c_{n',\pm m}\int_0^1 drr R^{\pm m}_{n'}\left(\frac{a_{\mathrm{fit}}}{a_w}r\right) R^{\pm m}_{n}\left(r\right) \\
&= 2\left(n+1\right)\sum_{n'=0}^{N} c_{n',\pm m}\sum_{k,k'=0}^{(n-m)/2}\left(\frac{a_{\mathrm{fit}}}{a_w}\right)^{n-2k} A_{n',m,k'}A_{n,m,k}\int_0^1 drr^{n'-2k'+n-2k+1} \\
&= 2\left(n+1\right)\sum_{n'=0}^{N} c_{n',\pm m}\sum_{k,k'=0}^{(n-m)/2}\left(\frac{a_{\mathrm{fit}}}{a_w}\right)^{n-2k} \frac{A_{n',m,k'}A_{n,m,k}}{2\left(n'+n-k'-k+1\right)} \qquad (10)
\end{aligned}
$$

for $n-m=$ even.

Figure 1 shows the result of evaluating Eq (10) for the spherical terms ($m=0$) with $n=n'$ ranging from 0 to 40 for 1 % error or uncertainty in the value of $a_w$, e.g. $\left|a_w-a_{\mathrm{fit}}\right|/a_{\mathrm{fit}}\ \sim 0.01$, also for an 0.1 % error. For the $\pm1$ % case, the error in the fitted coefficient is roughly equal, in percent, to the Zernike order itself.

Next, consider orthogonality. Sticking with the spherical terms, suppose that $c_{n',0}\neq 0$ only for one particular value of $n'$.

$$
c^{\mathrm{fit}}_{n,0} = 2\left(n+1\right)c_{n',0}\sum_{k,k'=0}^{n/2}\left(\frac{a_{\mathrm{fit}}}{a_w}\right)^{n-2k}\frac{A_{n',0,k'}A_{n,0,k}}{2\left(n'+n-k'-k+1\right)} \qquad (11)
$$

If $a_{\mathrm{fit}}=a_w$ then, since

$$
\sum_{k,k'=0}^{n/2}\frac{A_{n',0,k'}A_{n,0,k}}{2\left(n'+n-k'-k+1\right)} = \frac{1}{2\left(n+1\right)}\delta_{n,n'} \qquad (12)
$$

we have $c^{\mathrm{fit}}_{n',0}=c_{n',0}$ with all other $c^{\mathrm{fit}}_{n,0}=0$ and orthogonality is maintained. But if $a_{\mathrm{fit}}\neq a_w$ then

$$
\sum_{k,k'=0}^{n/2}\left(\frac{a_{\mathrm{fit}}}{a_w}\right)^{n-2k}\frac{A_{n',0,k'}A_{n,0,k}}{2\left(n'+n-k'-k+1\right)} \neq \frac{1}{2\left(n+1\right)}\delta_{n,n'}
$$

and orthogonality is broken. Figures 2 and 3 show this explicitly. Figure 2(3) shows the overlap integral of $Z^0_{10}\left(Z^0_{20}\right)$ with $Z^0_n$ for $n=0$ to 40 for a +1 % (squares) and $-1$ % (dots) error in the radius of the pupil used to do the fit. The red dot is the true overlap value of $Z^0_{10}\left(Z^0_{20}\right)$ with itself.

## V. CONCLUSION

The results shown in the Figures above illustrate the fact that in order to accurately fit a specific higher order Zernike coefficient to within a given accuracy, the uncertainty or error in the radius of the wavefront data used in the fit must get smaller and smaller as the order gets larger. For example, to fit coefficients up to the 40th order to within a few percent of the true value requires knowing the true value of the radius to on the order of $\pm0.1$ % or less. To put it differently, the above results indicate that, to get a reasonably accurate fit to high order Zernikes, the diameter of the wavefront data, as measured across a detector array, must be on the order of a thousand pixels or more.

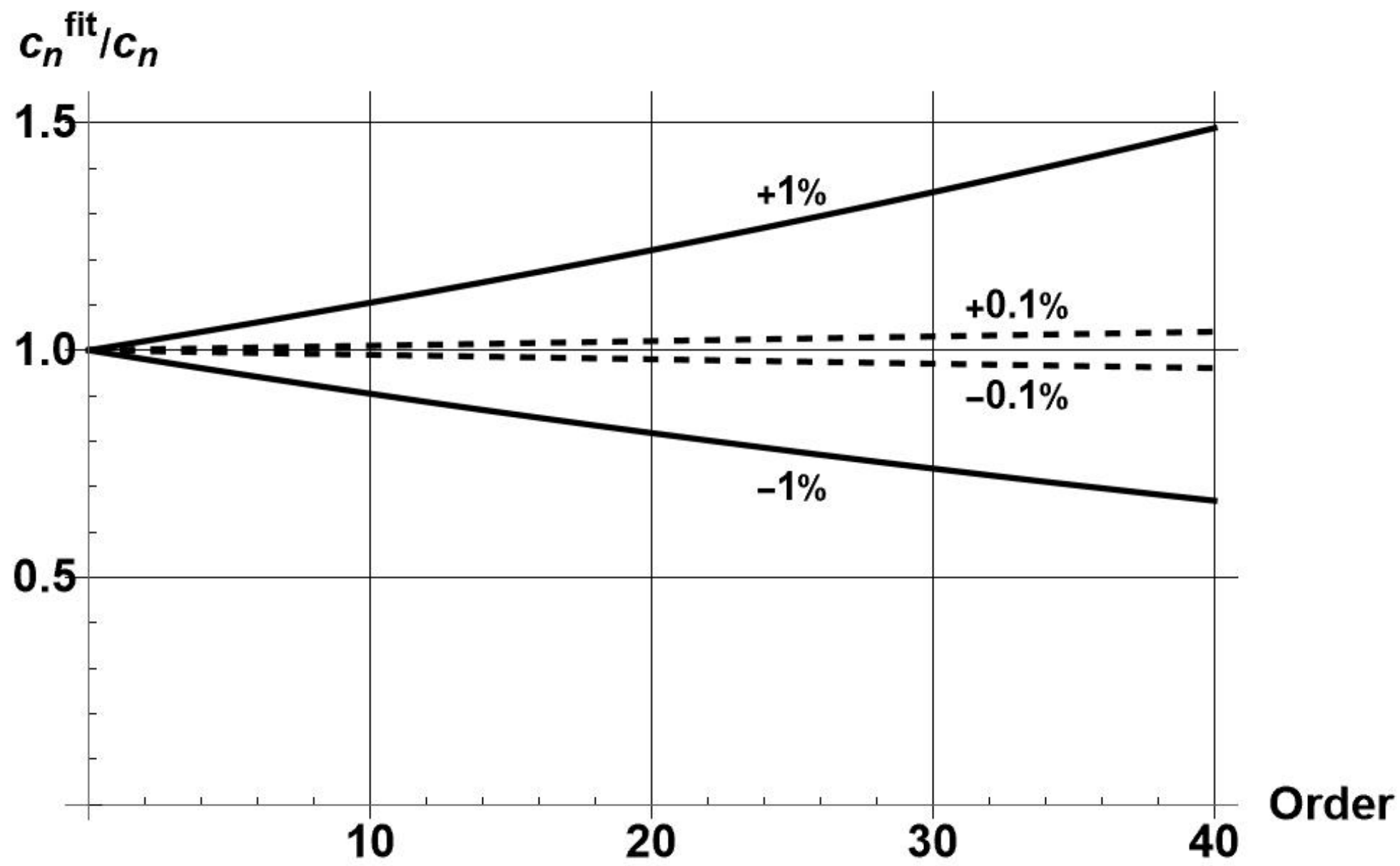

FIG. 1 Ratio of the fitted Zernike coefficient value to the actual coefficient value as a function of $n$, the order of the Zernike, for $\pm 1$ % and $\pm 0.1$ % error in the pupil radius used in the fit. Obviously for the fitted values to be within a few percent of the true value the error in the radius must be on the order $\pm 0.1$ % or less.

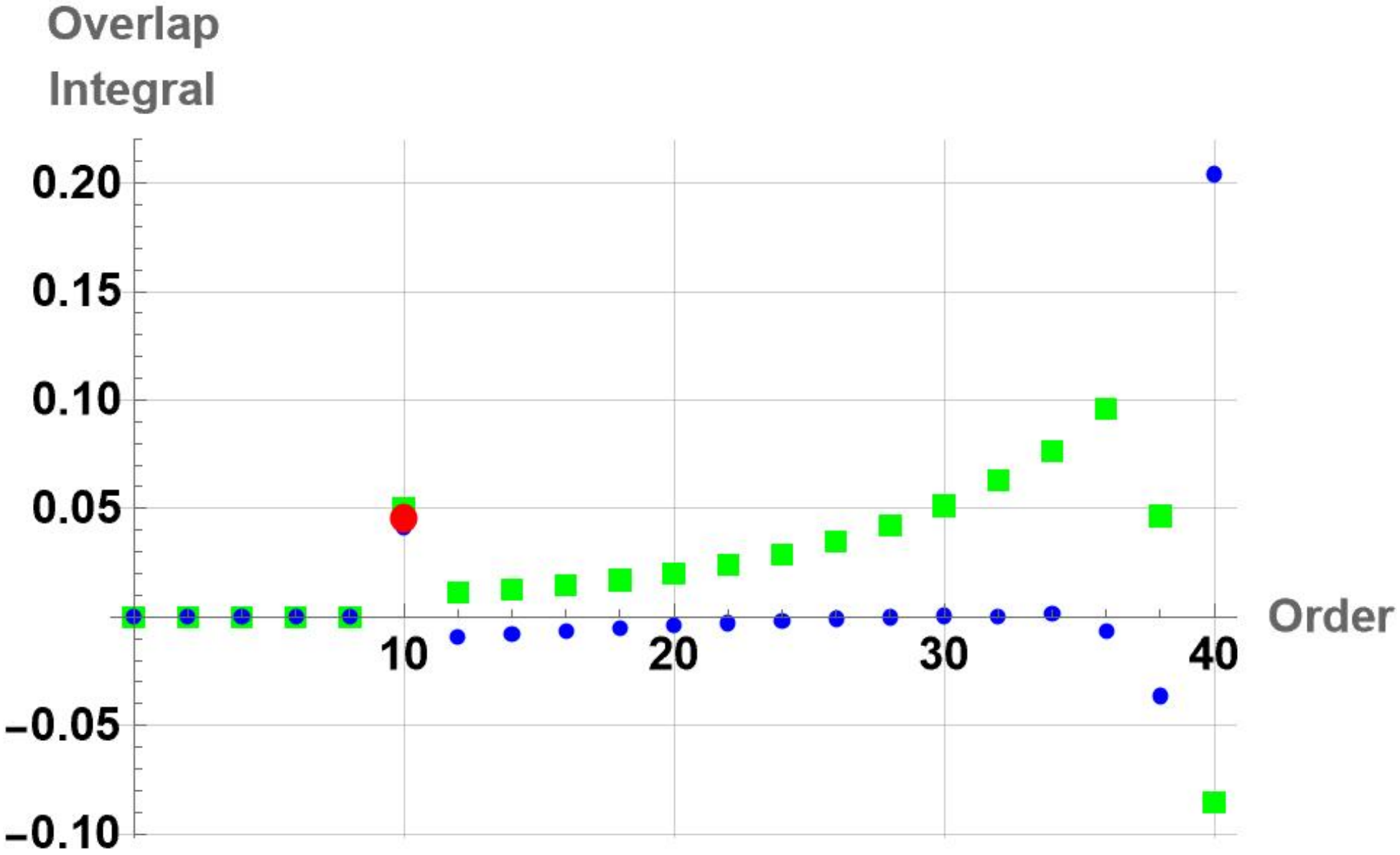

FIG. 2 Overlap integral of $Z_{10}^0$ with $Z_n^0$ for $n = 0$ to 40 for $+1$ % (Green Squares) and $-1$ % (Blue Dots) error in in the radius of the pupil used to do the fitting. The Red Dot is the true overlap value of $Z_{10}^0$ with itself, 1/22.

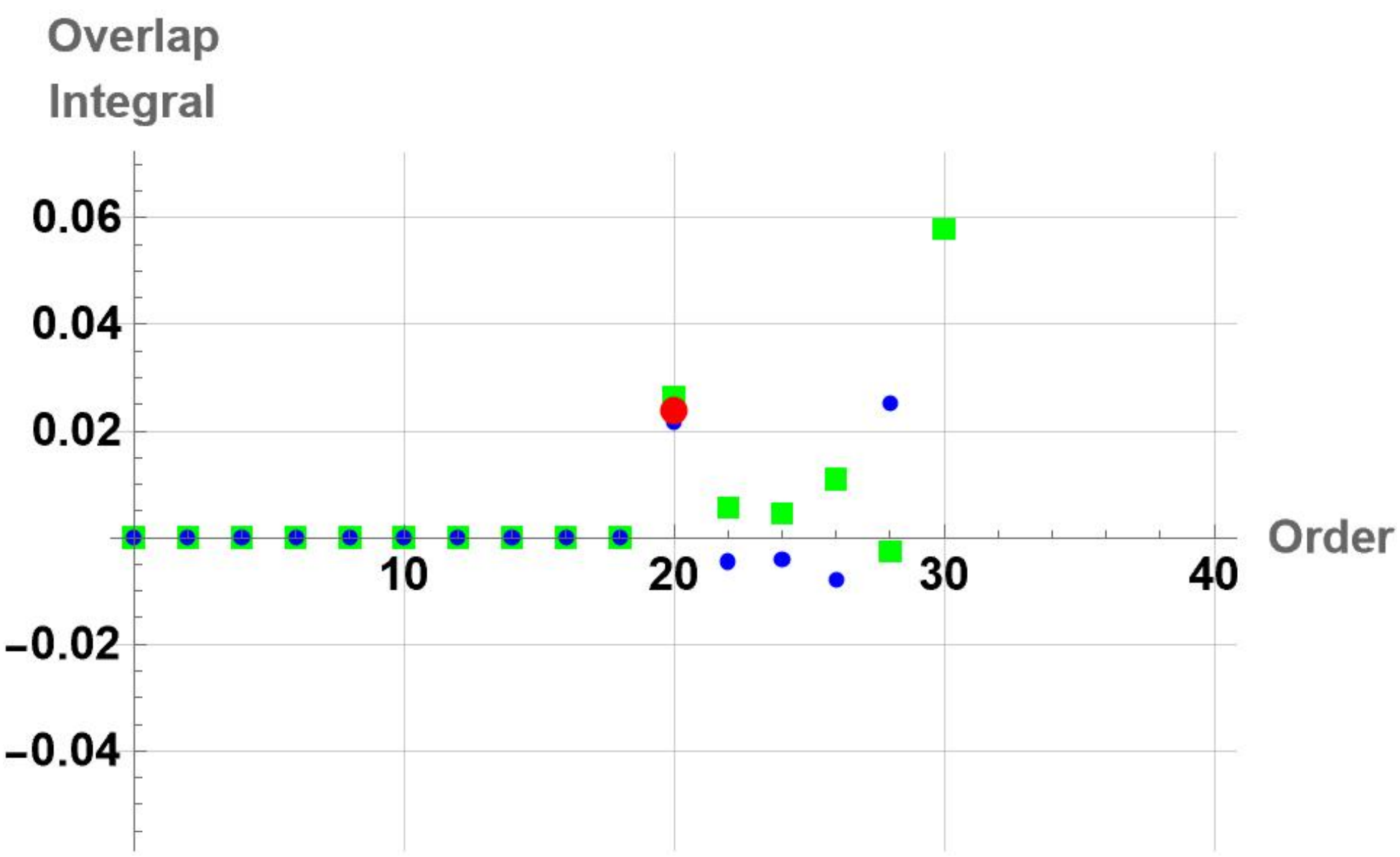

FIG. 3 Overlap integral of $Z_{20}^0$ with $Z_n^0$ for $n = 0$ to 40 for +1 % (Green Squares) and −1 % (Blue Dots) error in in the radius of the pupil used to do the fitting. The Red Dot is the true overlap value of $Z_{20}^0$ with itself, 1/42. The points above $n = 30$ are out of range on this scale. For example, for $n = 40$, the +1 % value is −1350 and the −1 % value is −792.

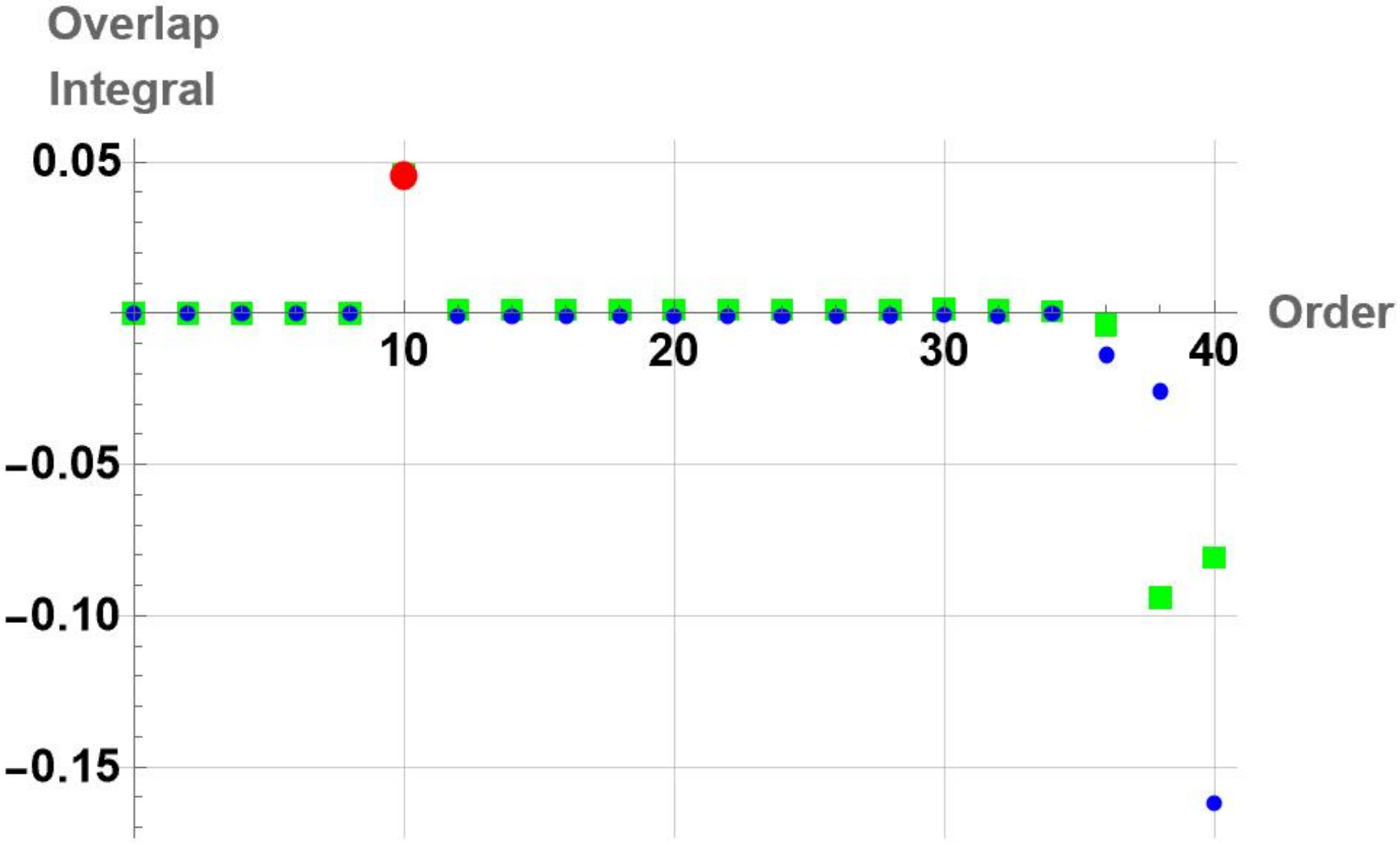

FIG. 4 Overlap integral of $Z_{10}^0$ with $Z_n^0$ for $n = 0$ to 40 for +0.1 % (Green Squares) and −0.1 % (Blue Dots) error in in the radius of the pupil used to do the fitting. The Red Dot is the true overlap value of $Z_{10}^0$ with itself, 1/22.

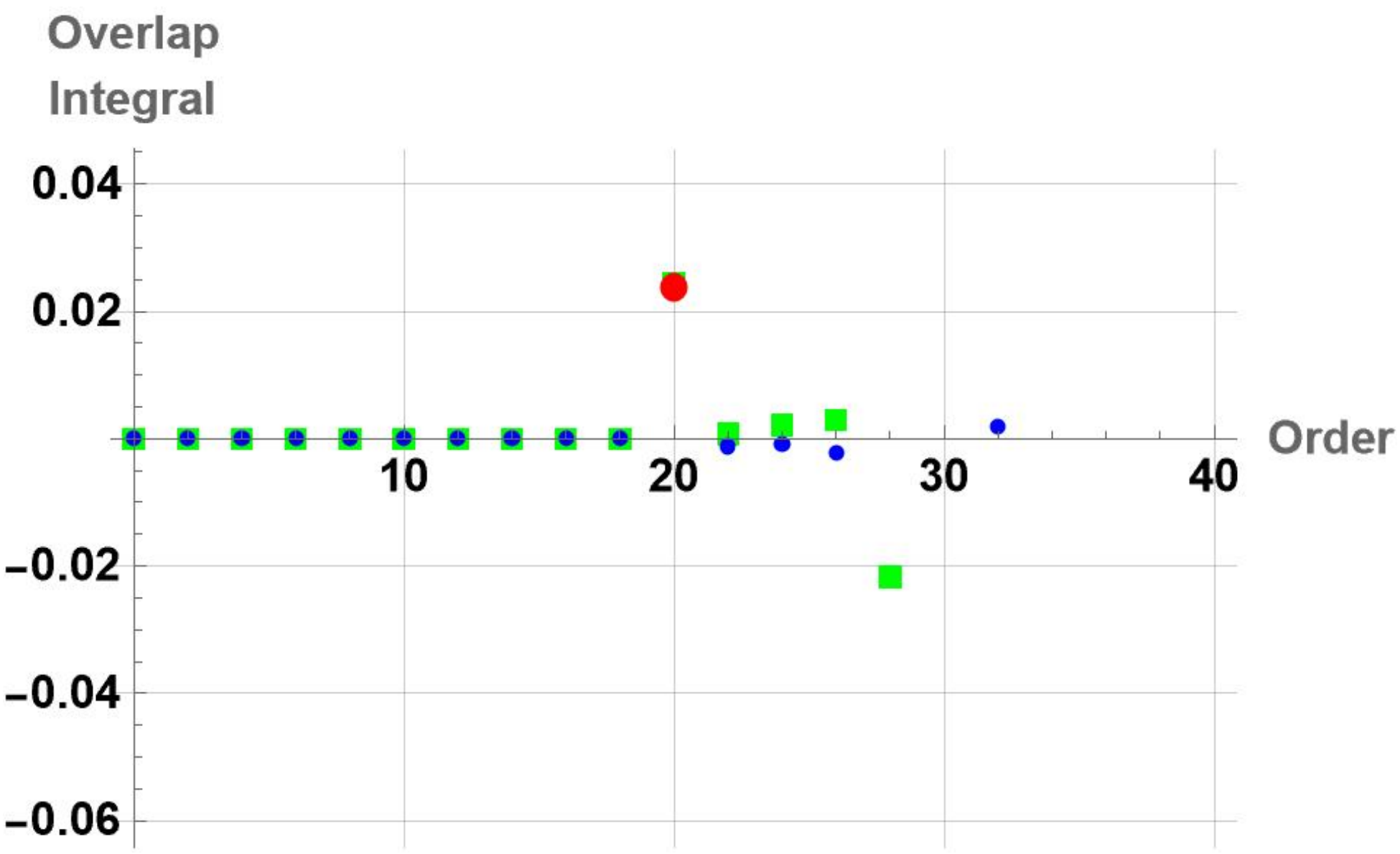

FIG. 5 Overlap integral of $Z^0_{20}$ with $Z^0_n$ for $n = 0$ to 40 for +0.1 % (Green Squares) and −0.1 % (Blue Dots) error in in the radius of the pupil used to do the fitting. The Red Dot is the true overlap value of $Z^0_{20}$ with itself, 1/42. The points above $n \simeq 30$ are out of range on this scale. For example, for $n = 40$, the +0.1 % value is −3193 and the −0.1 % value is −1251.